\documentclass[aip,jmp,preprint,amsmath,amssymb,showpacs]{revtex4-1}
\usepackage{times}
\usepackage{color}
\usepackage{float}
\usepackage{CJK}
\usepackage{epsfig}
\usepackage{amsmath}
\usepackage{lipsum}
\begin{document}
\begin{CJK*}{GBK}{}

\title{Double transition of information spreading in a two-layered network}
\author{Jiao Wu}
\affiliation{Business School, University of Shanghai for Science and Technology, Shanghai 200093, China}
\author{Muhua Zheng}
\affiliation{Departament de F\'{\i}sica de la Mat\`{e}ria Condensada, Universitat de Barcelona, Barcelona, Spain}
\affiliation{Universitat de Barcelona Institute of Complex Systems (UBICS), Universitat de Barcelona, Barcelona, Spain}
\author{Wei Wang}
\affiliation{Cybersecurity Research Institute, Sichuan University, Chengdu 610065, China}
\author{Huijie Yang}
\affiliation{Business School, University of Shanghai for Science and Technology, Shanghai 200093, China}
\author{Changgui Gu}
\email{gu\_changgui@163.com}
\affiliation{Business School, University of Shanghai for Science and Technology, Shanghai 200093, China}

\begin{abstract}
A great deal of significant progress has been seen in the study of information spreading on
populations of networked individuals. A common point in many of past studies is that there is only
one transition in the phase diagram of the final accepted size versus the transmission
probability. However, whether other factors alter this phenomenology is still under
debate, especially for the case of information spreading through many channels and platforms.
In the present study, we adopt a two-layered network to represent the
interactions of multiple channels and propose a SAR
(Susceptible-Accepted-Recovered) information spreading model. Interestingly, our model shows
a novel double transition including a continuous transition and a following discontinuous
transition in the phase diagram, which originates from two outbreaks between the two layers of the
network. Further, we reveal that the key factors are a weak coupling condition between the two layers, a large adoption
threshold and the difference of the degree distributions between the two layers. Then, an edge-based compartmental
theory is developed which fully explains all numerical results. Our findings may be of significance
for understanding the secondary outbreaks of the information in real life.
\end{abstract}
\maketitle

\begin{quotation}
In our life, with the fast development of the modern communication tools, people usually receive
the information from multiple channels, such as face-to-face interactions, telephone, live chat,
Facebook, Twitter and so on. Consequently, some new forms of information spreading have emerged
from one geographical region to another. Thus, how to understand these new communication styles affecting the
information spreading is a new challenging problem in network science. We here adopt a
two-layered network to represent the interactions of multiple channels and propose a SAR
(Susceptible-Accepted-Recovered) information spreading model. Our numerical simulations reveal
that, contrary to previous work, there is a double
transition, including a continuous transition and a following discontinuous transition in the
final accepted size with respect to a transmission probability. Further, we demonstrate that the
phenomenon of the double transition originates from two outbreaks in the two networks,
which depends on a weak coupling condition between the two networks, the difference of the degree
distributions between them, and a large adoption threshold in turn. Moreover, an edge-based compartmental theory is
developed which perfectly agree with the numerical simulations. These findings may enrich our
understanding of information spreading dynamics, especially in the aspect of information secondary
outbreaks.
\end{quotation}

\section{Introduction}
The spreading process is currently one of the hottest topics in the field of complex networks, such
as the spreading of epidemic, opinion, rumor, new technologies and behaviors and so on. So far, a
great deal of significant progresses have been achieved including the infinitesimal threshold
\cite{Pastor-Satorras:2001,Boguna:2002,Ferreira:2012,Boguna:2013,Parshani:2010,Castellano:2010},
reaction-diffusion model \cite{Colizza:2007a,Colizza:2007b,Andrea:2008,Liu:2009},
temporal and/or multilayer networks
\cite{Boccaletti:2014,Feng:2015,Sahneh:2013,Wang:2013,Yagan:2013,Newman:2005,Marceau:2011,
Buono:2015,Buono:2014,Zhao:2014,Zheng:2017,Zheng:2018,Holme:2012,Perra:2012} etc (see the review Refs.
\cite{Pastor:2015,Barrat:2008,Dorogovtsev:2008,Wang:2017} for details). These models significantly
increase our understanding on epidemic/information spreading and are very useful for public health
authorities and relevant government departments to control the epidemic/information spreading.

A common point in all these contributions is that there is only one transition in the
spreading process where the spreading range will be approximately zero when the transmission
probability $\beta$ is less than a critical value $\beta_c$ and become nonzero when $\beta \geq
\beta_c$. Larger than the critical point $\beta_c$, the spreading range will be gradually increased
with the further increase of $\beta$. On the other hand, in recent years, a novel double
transition was observed on epidemic spreading process in some particular conditions, such as
the network with a very heterogeneous and
clustered structure \cite{Simon:2014,Bhat:2017}, epidemic spreading with an asymmetric interaction
\cite{Allard:2017} and contagion processes with heterogeneous adoptability \cite{Min:2017} etc.
The so-called double transition indicates that there are two critical
values $\beta_c^1$ and $\beta_c^2$ in the spreading process. The first transition is between
healthy and endemic phases, and the second transition is between two endemic phases with very
different internal organizations. For example, Ref\cite{Allard:2017} shows that with $\beta
<\beta_c^{1}$, all outbreaks are microscopic and quickly die out; with $\beta_c^1<\beta <
\beta_c^{2}$, they observed a macroscopic epidemic within the network of homosexual contacts
between males, with microscopic spillover into the rest of the population via bisexual males. While
$\beta > \beta_c^{2}$, they found a more classic epidemic scenario in the sense that it is of
macroscopic scale in most of the population.

Although some significant mechanisms of double transition have been uncovered in the previous
studies, many gaps in our knowledge remain in spreading dynamics. For example, this unique double
transition was observed only on epidemic spreading dynamics in some particular situations.
However, the study of double transition on information spreading process is neglected,
especially in the aspect of identifying the critical factors driving this phenomenon. As we know,
the information spreading carries its special features, which is different with epidemic spreading,
such as memory effects (i.e., previous contacts could impact the information spreading in current
time \cite{Dodds:2004,Lu:2011,Zheng:2013}) and non-redundant contacts (people usually do not transfer an
information item more than once to the same guy \cite{Wang:2015,Wu:2018}). In addition, information
spreading is affected by multiple channels from different types of contacts in different regions
\cite{Brummitt:2012,Lee:2014,Min:2016}. For instance, when choosing which products to buy, ideas to
accept, and behaviors to adopt, people are not only influenced by friends, colleagues and family in
the same region through face-to-face interactions, but also affected by distant relatives and
friends in another region through the telephone or Internet communication. In this sense, it is
very necessary to investigate the double transition in the information spreading dynamics
with the effects of multiple channels and memory of non-redundant information.

The effects of multiple channels on the spreading process have been widely investigated based on a
powerful analytical framework: multilayer or multiplex
networks\cite{Boccaletti:2014,Feng:2015,Sahneh:2013,Wang:2013,Yagan:2013,Newman:2005,Marceau:2011,
Buono:2015,Buono:2014,Zhao:2014,Zheng:2017,Zheng:2018}, where the intra-links and inter-links
represent the multiple social relations (channels) among individuals. So far, the majority of
researches about multilayer networks are mainly focused on how the one-to-one interconnections
influence the dynamic processes taking place on them
\cite{Funk:2010,Allard:2009,Son:2012,Sanz:2012,Souza:2009,Hackett:2016,Dickison:2012,Mendiola:2012}.
However, to the best of our knowledge, few researchers pay attention to the information spreading with
one-to-many interconnections, especially in the aspect of mathematical theory analysis. On the other hand,
despite many studies have revealed that the interaction strength between different networks, degree-degree correlation,
degree distribution and mean degree in each network play a critical role in the
relevant dynamic processes\cite{Funk:2010,Allard:2009,Son:2012,Sanz:2012,Souza:2009,Hackett:2016,Dickison:2012,Mendiola:2012},
how the properties of the multilayer network structures affect the double transition of information
spreading is still under debate in network science.

To fill these gaps, in this work, we propose a SAR (Susceptible-Accepted-Recovered) information
spreading model on multilayer networks, where we emphasize the effects of multiple channels,
memory and non-redundant contacts. Our numerical simulations reveal that, contrary to previous
work, there is a double transition including a continuous transition and a following
discontinuous transition in the final accepted size with respect to a transmission probability.
Further, we demonstrate that the phenomenon of the double transition originates from two
outbreaks between the two networks, which depends on a weak coupling condition between the two
networks, the difference of the degree distributions between them, and a large adoption threshold in
turn. To better understand the findings, an edge-based compartmental theory is developed which perfectly agree
with the numerical simulations.

The rest of this paper is organized as follows. In Sec. II, a Susceptible-Accepted-Recovered (SAR)
model on a two-layered network was proposed to describe the multiple channels information spreading.
In Sec. III, an edge-based compartmental compartmental theory is given in detail. In Sec. V,
simulation results are presented. Finally, in Sec. VI, the
conclusions and discussions are presented.

\section{The Susceptible-Accepted-Recovered model on a two-layered network}
\begin{figure}
\centering
\includegraphics[width=1\linewidth]{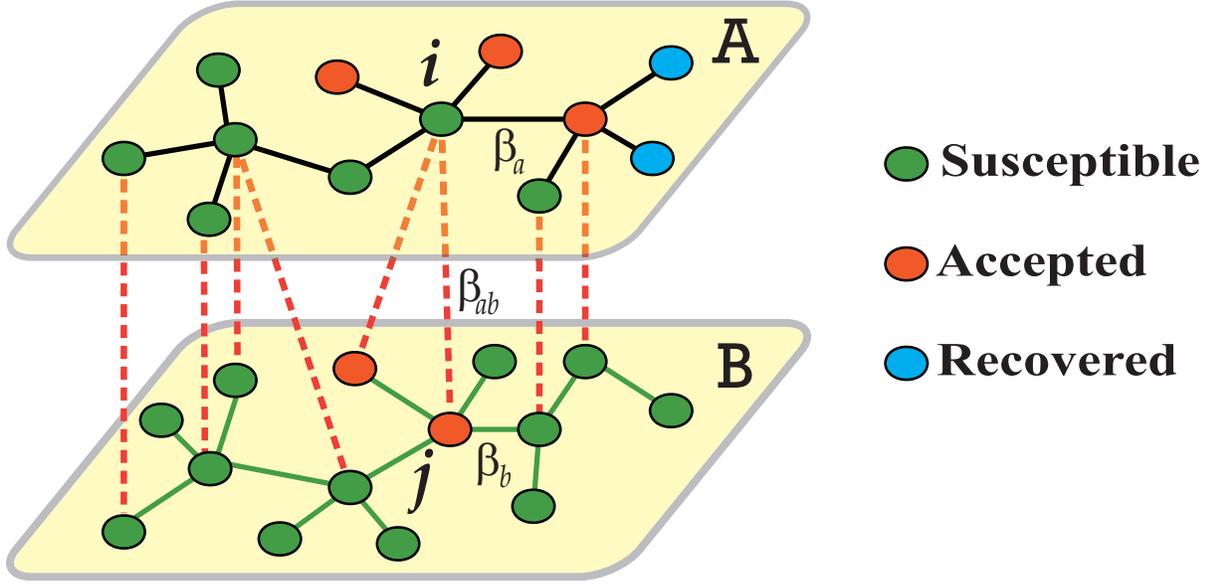}
\caption{(Color online). Sketch of the Susceptible-Accepted-Recovered (SAR) model on a two-layered
network. ``Black", ``green" and ``red" lines represent the links of the layer
$\mathcal{A}$, $\mathcal{B}$ and the inter-layer $\mathcal{AB}$, respectively. $\beta_a$, $\beta_b$ and
$\beta_{ab}$ denote the transmission probability of layers $\mathcal{A}$, $\mathcal{B}$ and
$\mathcal{AB}$. At time $t$, the susceptible node $i$ in layer $\mathcal{A}$ may receive a piece of
information from an accepted node in layer $\mathcal{A}$ and $\mathcal{B}$ with probability
$\beta_a$ and $\beta_{ab}$, respectively. Once the node $i$ receives the information successfully
from one accepted neighbor, the cumulative number $m$ of received information for node $i$ will
increase $1$ and the accepted neighbor will not transmit the same information to the node $i$ any
more. Assuming that the susceptible node $i$ has received the information $m$ times from the time
step $0$ to $t$, the node $i$ will become accepted state if $m\geq T_A$. }
\label{Fig:model}
\end{figure}

To understand the effects of multiple
channels in the information spreading process, we here introduce a two-layered network with
coupling between its two layers, i.e. the layer $\mathcal{A}$ and $\mathcal{B}$ in Fig.
\ref{Fig:model}. We let the two layers have the same size $N_a=N_b=N$ and their degree
distributions $P_A(k)$ and $P_B(k)$ be different. We may imagine the layer $\mathcal{A}$ as a
human communication network for one geographic region or community and the layer $\mathcal{B}$
for a separated region. There are two kinds of links for each node in the two-layered networks,
i.e. intra-links within layer $\mathcal{A}$ or $\mathcal{B}$ and the inter-links between layer $\mathcal{A}$
and $\mathcal{B}$. Each node could receive information not only through friends, colleagues and family
with intra-links in the same region, but also from distant relatives and friends with inter-links
in another region by the telephone and Internet. In details, we firstly generate two separated
networks $\mathcal{A}$ and $\mathcal{B}$ with the same size $N$ and different degree distributions
$P_A(k_a)$ and $P_B(k_b)$, respectively. Then, we add links randomly between $\mathcal{A}$ and
$\mathcal{B}$ until the steps we planned. The average node degrees of layer $\mathcal{A}$,
$\mathcal{B}$ and inter-layer $\mathcal{AB}$ is presented by $\langle k_a\rangle$,
$\langle k_b\rangle$, and $\langle k_{ab}\rangle$,
respectively. In the above way, we obtain an uncorrelated two-layered network.

To discuss information spreading in the two-layered network, we adopt a
Susceptible-Accepted-Recovered (SAR) model. At each time step, a node can occupy only one of the
three states: (i) Susceptible: the node has not received the information yet or has received the
information but hesitate to accept it; (ii) Accepted: the node accepts the information and
transmits it to its neighbors; (iii) Recovered: the node loses interest to the information and will
not spread it any more. Thus, this Susceptible-Accepted-Recovered (SAR) model is similar to the SIR
(Susceptible-Infected-Refractory) model in epidemiology.

The information spreading process can be described as follows:

(i) At the beginning, a fraction $\rho_0$ of nodes are random uniformly chosen from the layer $\mathcal{A}$ as
seeds (accepted state) to spread the first piece of information. All other nodes are in the susceptible state.

(ii) At each time step $t$, the susceptible node $i$ in layer $\mathcal{A}$ may receive a piece of
information from an accepted node in layer $\mathcal{A}$ and $\mathcal{B}$ with probability
$\beta_a$ and $\beta_{ab}$ (see Fig. \ref{Fig:model}), respectively. For the susceptible node in
layer $\mathcal{B}$, the change of the nodes' state is the same as in layer $\mathcal{A}$ but with probability $\beta_b$. Once
the node $i$ receives the information successfully from an accepted neighbor, the cumulative number
$m$ of received information for the node $i$ will increase one and this accepted neighbor will not
transmit the same information to the node $i$ any more, i.e., non-redundant information
transmission. As an individual has to remember the pieces of non-redundant information he or
she received from neighbors before time $t$, the so-called non-redundant information memory is
induced in our model.

(iii) When a susceptible node $i$ has received the information $m$ times until time step $t$
and $m\geq T_A$ in layer $\mathcal{A}$ (or $m\geq T_B$ in layer $\mathcal{B}$),
the node $i$ will become accepted state, where $T_A$ and $T_B$ is the adoption threshold of node
in layer $\mathcal{A}$ and $\mathcal{B}$, respectively. At the same time step, each accepted
node will lose interest in transmitting the
information and becomes recovered with probability $\mu$.

(iv) The steps are repeated until there is no accepted node in the network.

In our numerical simulations, we set the network size $N_a=N_b=10\,000$, recovered
probability $\mu=1.0$, $\beta_a=\beta_b=\beta$, and initially
chose $\rho_0=0.05$ of nodes in layer $\mathcal{A}$ to be accepted.

\section{theory}
\subsection{The edge-based compartmental theory on a single network}
Let us first illustrate the edge-based compartmental theory for a
single network, by following the methods introduced
in Refs. \cite{Wang:2015,Volz:2008,Miller:2011,Shu:2016,Miller:2012,Miller:2013a,Miller:2014}. We let $\rho_S(t)$,
$\rho_A(t)$, and $\rho_R(t)$ be the densities of the Susceptible, Accepted, and Recovered nodes at time $t$,
respectively. The spreading process will be ended when $t\rightarrow\infty$ and thus $\rho_R(\infty)$ represent the
final fraction of accepted nodes.

We use a variable $\theta(t)$ to denote the probability that a node $v$ has not transmitted the information to the
node $u$ along a randomly chosen edge by time $t$.
For an uncorrelated, large and sparse network, the probability
that a randomly chosen node $u$ of degree $k$ has received the information $m$ times from his/her neighbors at
time $t$ is
\begin{equation}\label{eq:1}
\tau(k,m,\theta(t))=(_{m}^{k})\theta(t)^{k-m}(1-\theta(t))^{m}
\end{equation}

Notice that a node with degree $k$ has the probability $1-\rho_0$ to be not one of the initial seeds. At the
same time, the probability that a susceptible node $u$ with degree $k$ has received the information $m$ times
and still does not accept it by time $t$ is $\sum_{m=0}^{T-1} \tau(k,m,\theta(t))$, where $T$ is the adoption threshold
in the model. Combining the initial seeds and summing over all possible values of $m$, we obtain the probability that the node
$u$ is still in the susceptible state at time $t$ as
\begin{equation}\label{eq:2}
s(k,t)=(1-\rho_0)\sum_{m=0}^{T-1} \tau(k,m,\theta(t))
\end{equation}

Averaging over all $k$, the density of susceptible nodes (i.e., the probability of a randomly chosen individual
is in the susceptible state) at time $t$ is given by
\begin{eqnarray}\label{eq:3}
\rho_S(t)=\sum\limits_{k=0}^{\infty}P(k)s(k,t).
\end{eqnarray}
where $P(k)$ is the degree distribution of the network. In order to solve $\rho_S(t)$, one needs to know $\theta(t)$.
Since a neighbor $v$ of node $u$ may be susceptible, accepted, or recovered, $\theta(t)$ can be expressed as
\begin{equation}\label{eq:4}
\theta(t)=\Phi^S(t)+\Phi^A(t)+\Phi^R(t)
\end{equation}
where $\Phi^S(t),\Phi^A(t),\Phi^R(t)$ is the probability that the neighbor $v$ is in the susceptible, accepted,
recovery state, respectively, and has not transmitted the information to node $u$ through this connection. Once
these three parameters derived, we will get the density of susceptible nodes at time $t$
by substituting them into Eq. (\ref{eq:1})(\ref{eq:2}) and then into Eq. (\ref{eq:3}). For this
purpose, in the following, we will focus on how to solve them.

To find $\Phi^S(t)$, we now consider a randomly chosen node $u$, and
assume this node is in the cavity state, which means that it cannot transmit any information to its neighbors $v$ but
can be informed by its neighbors. In this case, the neighbor $v$ can only get
the information from its other neighbors except the node $u$. If a neighboring node $v$ of $u$ has degree $k'$,
the probability that node $v$ has received $m$ pieces of the information at time $t$ will be $\tau(k'-1,m,\theta(t))=(_{m}^{k'-1})\theta(t)^{k'-1-m}(1-\theta(t))^{m}$.
After received the information $m$ times, node $v$ still does not accept it with probability
$(1-\rho_0)\sum_{m=0}^{T-1} \tau(k'-1,m,\theta(t))$. For uncorrelated networks,
the probability that one edge from node $u$ connects with a node $v$ with degree $k'$ is
$k'P(k')/\langle k \rangle$, where $\langle k\rangle$ is the mean degree of the network.
So, summing over all possible $k'$, one obtains
\begin{equation}\label{eq:5}
\Phi^S(t)=(1-\rho_0) \frac{\sum\limits_{k'}k'P(k')\sum\limits_{m=0}^{T-1}\tau(k'-1,m,\theta(t))}{\langle k \rangle}
\end{equation}

The growth of $\Phi^R(t)$ includes two
consecutive events: firstly, an accepted neighbor has not transmitted the information successfully to node $u$
with probability $1-\beta$; secondly, the accepted neighbor has become recovered with probability $\mu$. Combining
these two events, the $\Phi^A(t)$ to $\Phi^R(t)$ flux is $\mu(1-\beta)\Phi^A(t)$. Thus, one gets
\begin{equation}\label{eq:6}
\frac{d\Phi^R(t)}{dt}= \mu(1-\beta)\Phi^A(t)
\end{equation}

Once the accepted neighbor $v$ transmits the information to $u$ successfully (with probability $\beta$), the
$\Phi^A(t)$ to $1-\theta(t)$ flux will be $\beta\Phi^A(t)$, which means
\begin{eqnarray}\label{eq:7}
\frac{d(1-\theta(t))}{dt}=\beta\Phi^A(t).
\end{eqnarray}
That is
\begin{equation}\label{eq:8}
\frac{d\theta(t)}{dt}=-\beta\Phi^A(t).
\end{equation}

Combining Eqs. (\ref{eq:6}) and (\ref{eq:8}) and considering (as initial conditions) $\theta(0)=1$ and
$\Phi^R(0)=0$, one obtains
\begin{eqnarray}\label{eq:9}
\Phi^R(t)=\frac{\mu[1-\theta(t)](1-\beta)}{\beta}.
\end{eqnarray}

Substituting Eqs. (\ref{eq:5}) and (\ref{eq:9}) into Eq.(\ref{eq:4}), we get an expression for $\Phi^A(t)$ in terms
of $\theta(t)$. Then, one can rewrite Eq. (\ref{eq:8}) as
\begin{eqnarray}
\frac{d\theta(t)}{dt}&=&-\beta\theta(t)+\mu(1-\theta(t))(1-\beta)\nonumber \\
&&+\frac{\beta(1-\rho_0)\sum_{k'}k'P(k')\sum\limits_{m=0}^{T-1}\tau(k'-1,m,\theta(t))}{\langle k\rangle} \label{eq:10}
\end{eqnarray}

With $\theta(t)$ on hand, the equation of the system comes out to be
\begin{eqnarray}\label{eq:11}
\frac{d\rho_R(t)}{dt}&=&\mu \rho_A(t) \nonumber, \\
\rho_S(t)&=&\sum\limits_{k=0}^{\infty}P(k)s(k,t) \nonumber, \\
\rho_A(t)&=&1-\rho_S(t)-\rho_R(t).
\end{eqnarray}

In fact, Eq. (\ref{eq:10}) does not depend on Eq. (\ref{eq:11}), so the system is
governed by the single ordinary differential equation (\ref{eq:10}).
Although the resulting equation are simpler than those found by other methods,
it can be proven to exactly predict the disease/information spreading dynamics in the large-population
limit for different network topologies\cite{Wang:2015,Volz:2008,Miller:2011,Shu:2016,Miller:2012,Miller:2013a,Miller:2014}.

\subsection{The edge-based compartmental theory on a two-layered network}

Now, we develop an analogous theoretical framework from the single network to the case of two uncorrelated interconnected networks
based on the approach in Refs. \cite{Zheng:2018}, which is suited to the problems studied in
our work. In particular, when one assumes that the population is made up of two layers,
then $P_j(k_1,k_2)$ denote the probability
that a node of layer $j$ has $k_1$ degree in layer $1$ and $k_2$ in layer $2$.
For the sake of simplicity, one can name the two
layers $\mathcal{A}$ and $\mathcal{B}$ as $1$ and $2$. Let $\beta_{j,l}$ be the rate of
transmission across an edge from network $l$ to network $j$, and let us define $\mu$ to be the recovery rate of a node
in any layer.

Firstly, let us define $\theta_{j,l}$ to be the probability that randomly chosen an edge $(u,v)$, node $v$ in
layer $j$ ($j=1,2$) has not transmitted the information to the node $u$ in layer $l$ ($l=1,2$) by time $t$.
For the considered case, we have $\theta_{1,2}$, $\theta_{1,1}$, $\theta_{2,1}$ and $\theta_{2,2}$ four variables.
Once the four variables were obtained, we can solve the equations of the system.

Now, we will solve $\theta_{1,2}$ as an example in detail. Similarly to the the case of single network,
a neighbor $v$ in layer 2 of node $u$ in layer 1 may be susceptible, accepted, or recovered. Then $\theta_{1,2}$
can be expressed as \begin{equation}\label{eq:12}
\theta_{1,2}=\Phi^S_{1,2}+\Phi^A_{1,2}+\Phi^R_{1,2}
\end{equation}
where $\Phi^S_{1,2}$, $\Phi^A_{1,2}$, $\Phi^R_{1,2}$ is the probability that the neighbor $v$ is in the
susceptible, accepted, recovery state, and has not transmitted the information to node $u$ through this edge $(u,v)$.

Similarly, to find $\Phi^S_{1,2}$, the neighbor $v$ in layer 2 can only get
the information from its other neighbors except the node $u$ in layer 1. Thus,
the probability that the node $v$ with degree $(k_1,k_2)$ has received the information $m$ times from his/her neighbors at
time $t$ is $\tau(k_1-1,n,\theta_{2,1})\tau(k_2,m-n,\theta_{2,2})$, where $\tau(k_1-1,n,\theta_{2,1})$ indicates the probability
that the node $v$ received $n$ times information from $k_1-1$ neighbors with $\theta_{2,1}$ and $\tau(k_2,m-n,\theta_{2,2})$ is the
probability that the node $v$ received the last $m-n$ times information from $k_2$ neighbors with $\theta_{2,2}$. It should be
noted that function $\tau(k,m,\theta)=(_{m}^{k})\theta^{k-m}(1-\theta)^{m}$, which has the similar expression as Eq. (\ref{eq:1}).
After received the information $m$ times, node $v$ still does not accept it with probability
\begin{eqnarray}
X_{1,2}=\sum\limits_{m=0}^{T_B-1}\sum\limits_{n=0}^{m}
\tau(k_1-1,n,\theta_{2,1})\tau(k_2,m-n,\theta_{2,2}) \label{eq:13}
\end{eqnarray}
For uncorrelated networks, the probability that one
edge from node $u$ connects with a node $v$ with degree $(k_1,k_2)$ is
$\frac{k_1P_2(k_1,k_2)}{\sum_{k_1,k_2}k_1P_2(k_1,k_2)}$. Thus, one has

\begin{eqnarray}
\Phi^S_{1,2}&=& \frac{\sum\limits_{k_1,k_2}k_1P_2(k_1,k_2)X_{1,2}}{\sum\limits_{k_1,k_2}k_1P_2(k_1,k_2)} \label{eq:14}
\end{eqnarray}

It is easily to know that the growth of $\Phi^R_{1,2}$ includes two consecutive events:
first, an accepted neighbor has not transmitted the information to
node $u$ via with probability $1-\theta_{1,2}$; second, the accepted neighbor has been recovered with
probability $\mu$. Combining these two events, the $\Phi^A_{1,2}$ to $\Phi^R_{1,2}$ flux is $\mu(1-\theta_{1,2})\Phi^A_{1,2}$.
Thus, one gets
\begin{eqnarray}
\frac{d\Phi^R_{1,2}}{dt}&=& \mu(1-\theta_{1,2})\Phi^A_{1,2} \label{eq:15}
\end{eqnarray}

Once the accepted neighbor $v$ in layer 2 transmits the information to node $u$ in layer 1 successfully (with
probability $\beta_{1,2}$), the $\Phi^A_{1,2}$ to $1-\theta_{1,2}$ flux will be $\beta_{1,2}\Phi^A_{1,2}$, which means
\begin{eqnarray}
\frac{d\theta_{1,2}}{dt}&=&-\beta_{1,2}\Phi^A_{1,2} \label{eq:16}
\end{eqnarray}

Combining Eqs. (\ref{eq:15}) and (\ref{eq:16}), and considering the initial conditions $\theta_{1,2}(0)=1$ and
$\Phi^R_{1,2}(0)=0$, one obtains
\begin{eqnarray}
\Phi^R_{1,2}&=&\frac{\mu(1-\theta_{1,2})(1-\beta_{1,2})}{\beta_{1,2}} \label{eq:17}
\end{eqnarray}
Substituting Eqs. (\ref{eq:14}) (\ref{eq:17}) into Eq.(\ref{eq:12}) and then into (\ref{eq:16}) , one gets

\begin{eqnarray}\label{eq:18}
\dot{\theta}_{1,2} &=& -\beta_{1,2}(\theta_{1,2}-\Phi^S_{1,2}-\Phi^R_{1,2})\nonumber\\
&=&-\beta_{1,2}\theta_{1,2}+\mu(1-\theta_{1,2})(1-\beta_{1,2})\nonumber\\
&&+\beta_{1,2} \frac{\sum\limits_{k_1,k_2}k_1P_2(k_1,k_2)X_{1,2}}{\sum\limits_{k_1,k_2}k_1P_2(k_1,k_2)}
\end{eqnarray}

Similarly, one can write down $\theta_{1,1}$, $\theta_{2,1}$ and $\theta_{2,2}$ as follows
\begin{eqnarray}
\dot{\theta}_{1,1} &=& -\beta_{1,1}(\theta_{1,1}-\Phi^S_{1,1}-\Phi^R_{1,1})\nonumber\\
&=& -\beta_{1,1}\theta_{1,1}+\mu(1-\theta_{1,1})(1-\beta_{1,1}) \nonumber\\
&&+\beta_{1,1} \frac{(1-\rho_0)\sum\limits_{k_1,k_2}k_1P_1(k_1,k_2)X_{1,1}}{\sum\limits_{k_1,k_2}k_1P_1(k_1,k_2)} \label{eq:19} \\
\dot{\theta}_{2,1}&=&\!-\beta_{2,1}(\theta_{2,1}-\Phi^S_{2,1}-\Phi^R_{2,1})\nonumber\\
&=&-\beta_{2,1}\theta_{2,1} +\mu(1-\theta_{2,1})(1-\beta_{2,1}) \nonumber\\
&&+ \beta_{2,1} \frac{(1-\rho_0)\sum\limits_{k_1,k_2}k_2P_1(k_1,k_2)X_{2,1}}{\sum\limits_{k_1,k_2}k_2P_1(k_1,k_2)}\label{eq:20}\\
\dot{\theta}_{2,2}&=&-\beta_{2,2}(\theta_{2,2}-\Phi^S_{2,2}-\Phi^R_{2,2})\nonumber\\
&=&-\beta_{2,2}\theta_{2,2} +\mu(1-\theta_{2,2})(1-\beta_{2,2})\nonumber\\
&&+\beta_{2,2} \frac{\sum\limits_{k_1,k_2}k_2P_2(k_1,k_2)X_{2,2}}{\sum\limits_{k_1,k_2}k_2P_2(k_1,k_2)}\label{eq:21}
\end{eqnarray}
where
\begin{eqnarray}
X_{1,1}&=&\sum\limits_{m=0}^{T_A-1}\sum\limits_{n=0}^{m}\tau(k_1-1,n,\theta_{1,1})\tau(k_2,m-n,\theta_{1,2}) \label{eq:22} \\
X_{2,1}&=&\sum\limits_{m=0}^{T_A-1}\sum\limits_{n=0}^{m}\tau(k_1,n,\theta_{1,1})\tau(k_2-1,m-n,\theta_{1,2}) \label{eq:23} \\
X_{2,2}&=&\sum\limits_{m=0}^{T_B-1}\sum\limits_{n=0}^{m}\tau(k_1,n,\theta_{2,1})\tau(k_2-1,m-n,\theta_{2,2}) \label{eq:24}
\end{eqnarray}
It should be noted that as a node in layer $1$ has the probability $1-\rho_0$ not to be one of the initial seeds,
after received the information $m$ times, node $v$ through a corresponding edge still does not accept it with probability
$(1-\rho_0)X_{1,1}$ and $(1-\rho_0)X_{2,1}$ in Eqs. (\ref{eq:19}) and (\ref{eq:20}), respectively. With Eqs. (\ref{eq:18}-\ref{eq:24}) on hand,
the densities associated with each distinct state can be obtained by
\begin{equation}
\begin{cases}
\dot{R}_1=\mu A_1(t) \\
S_1(t)=(1-\rho_0)\sum\limits_{k_1,k_2}^\infty P_1(k_1,k_2)Y_1\\ \label{eq:25}
A_1(t)=1-S_1(t)-R_1(t)
\end{cases}
\end{equation}

\begin{equation}
\begin{cases}
\dot{R}_2=\mu A_2(t) \\
S_2(t)=\sum\limits_{k_1,k_2}^\infty P_2(k_1,k_2)Y_2\\ \label{eq:26}
A_2(t)= 1-S_2(t)-R_2(t)
\end{cases}
\end{equation}
where
\begin{eqnarray}
Y_1=\sum\limits_{m=0}^{T_A-1}\sum\limits_{n=0}^{m}\tau(k_1,n,\theta_{1,1})\tau(k_2,m-n,\theta_{1,2})\label{eq:27}\\
Y_2=\sum\limits_{m=0}^{T_B-1}\sum\limits_{n=0}^{m}\tau(k_1,n,\theta_{2,1})\tau(k_2,m-n,\theta_{2,2})\label{eq:28}
\end{eqnarray}
Eqs. (\ref{eq:25}) and (\ref{eq:26}) are the main theoretical results in this paper. To obtain the densities
associated with each state, instead of getting the analytic solutions of Eqs. (\ref{eq:25}) and (\ref{eq:26}),
we solve them by numerical integration and get the corresponding theoretical curves.

\section{Results}
To study the effects of multiple channels on information spreading, we have performed extensive
simulations with our model in coupled Scale-free (SF)\cite{Catanzaro:2005}
and Erdos-R\H{e}nyi (ER) networks \cite{Albert:2002}. To compare the theoretical predictions with the
numerical results, we also take into account coupled ER-ER and SF-SF networks in this work. Next,
we mainly try to find out the key factors, which influence
the emergence of the double transition on information spreading process.
\subsection{The effects of multiple channels on the double transition}
\begin{figure}
\centering
\includegraphics[width=0.7\linewidth]{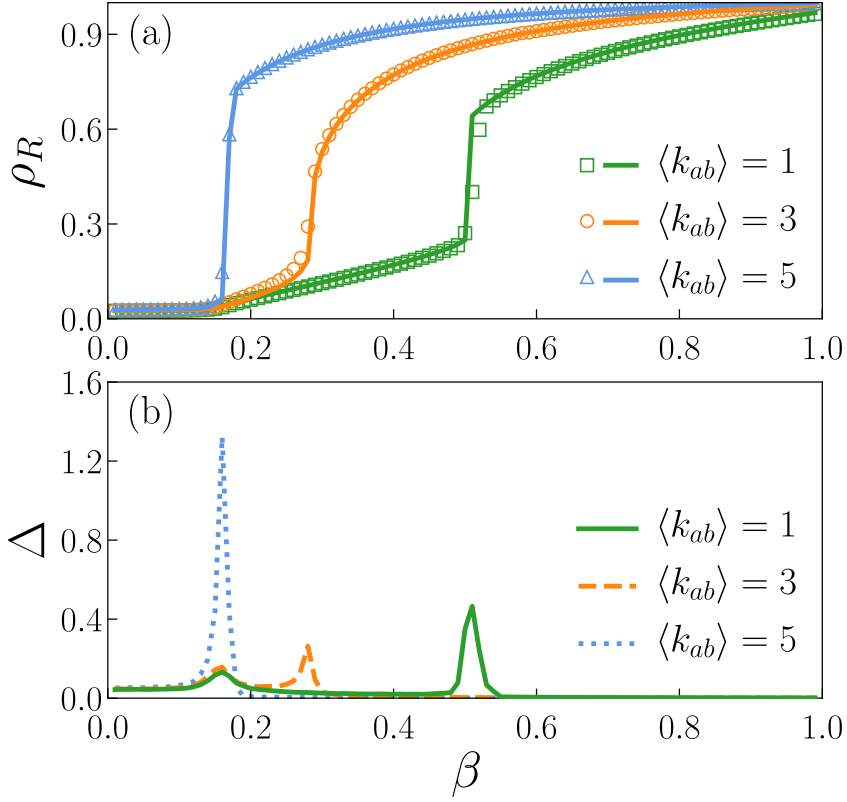}
\caption{(Color online). Emergence of the double transition on information spreading process
on SF-ER networks. (a) and (b) represent the density of final recovered nodes $\rho_R$ and
variability $\Delta$ as a function of transmission probability $\beta$ with different average
degree $\langle k_{ab}\rangle$, respectively. Squares, circles and up triangles represent $\langle k_{ab}\rangle=1$, $3$ and $5$,
respectively. The symbols show the simulated results and the lines are the corresponding
theoretical results in (a) from Eqs. (\ref{eq:25}) and (\ref{eq:26}). The results
are averaged over $10^3$ independent realizations. The parameters are $N_a=N_b=10\,000$, $\mu=1.0$,
$\beta_{ab}=0.5$, $T_A=T_B=2$, $\rho_0=0.05$, $P_A(k_a)\sim k_a^{-2.1}$, $\langle k_a\rangle=6$, $\langle k_b\rangle=6$. }
\label{Fig:kab}
\end{figure}

To better quantify the spreading behavior, we let $\rho_S(t)$, $\rho_A(t)$ and $\rho_R(t)$ denote
the fraction of susceptible, accepted and recovered nodes at time $t$ in the whole network. When the spreading
is ended, the final size of recovered nodes can be denoted by $\rho_R$. A
larger $\rho_R$ implies a larger spreading range at the final state.
To numerically identify the effective spreading threshold $\beta_c$ of
the SAR model, we use the variability measure\cite{Shu:2016,Crepey:2015}:
\begin{eqnarray}
\Delta=\frac{\sqrt{\langle \rho_R^2\rangle-\langle \rho_R\rangle^2}}{\langle \rho_R\rangle}
\label{eq:23}
\end{eqnarray}
In general, the variability $\Delta$ exhibits a peak at a critical
point\cite{Shu:2016,Crepey:2015}. Thus, we estimate the numerical effective spreading threshold
$\beta_c$ from the position of the peak of the variability.

Fig. \ref{Fig:kab}(a) shows the final size of recovered nodes $\rho_R$ as a function of
transmission probability $\beta$ with different average degree $\langle k_{ab}\rangle$ on SF-ER networks. Fig.
\ref{Fig:kab}(b) shows the variability $\Delta$ versus $\beta$ with corresponding $\langle k_{ab}\rangle$ in Fig.
\ref{Fig:kab}(a). When the interaction strength is weak (i.e., $\langle k_{ab}\rangle$ is relatively small), the
double transition occur on the information spreading process, which is indicated by two peaks of
$\Delta$ in Fig. \ref{Fig:kab}(b). It has also found that the system undergoes a
continuous transition from accepted free phase to accepted phase and a following discontinuous
transition between the accepted phases. In addition, with the increasing of $\langle k_{ab}\rangle$,
the second critical point $\beta_c^2$ close to the
first one $\beta_c^1$. Once the coupling strength is strong enough ($\langle k_{ab}\rangle=5$), the two
critical points merge into one, i.e., the second transition is vanished. These result have
been confirmed by Eqs. (\ref{eq:25}) and (\ref{eq:26}) of the theory, see the lines in Fig.
\ref{Fig:kab}(a). It is maybe helpful to understand the influence of $\langle k_{ab}\rangle$ on the double
transition from the aspect of purely coupling in network structure. When the coupling strength is
strong, a two-layered network behave as a solid single network \cite{Radicchi:2013,Sahneh:2015}. In
this case, the effect of multiple channels is not prominent and the spreading behavior is the same as
the common one \cite{Wang:2015}. Therefore, a key factor determining the occurrence of double
transition is a weak coupling between two networks.

\begin{figure}
\centering
\includegraphics[width=0.8\linewidth]{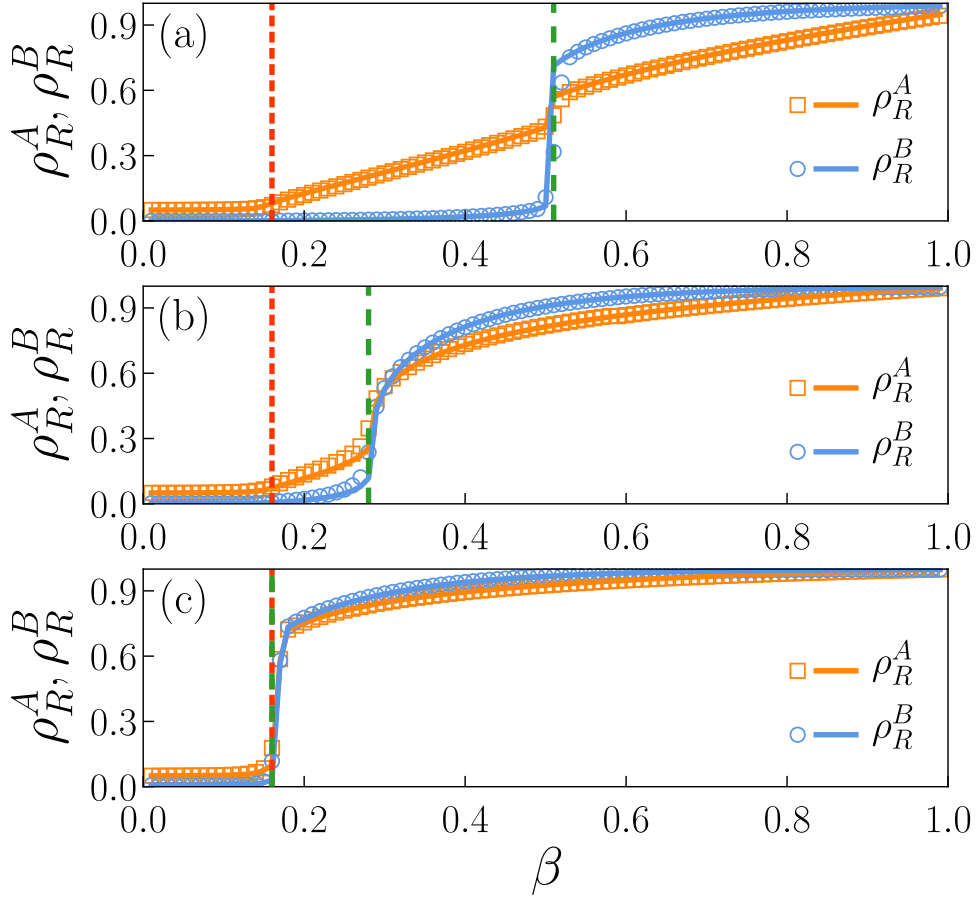}
\caption{(Color online). The densities of final recovered nodes $\rho_R^A$ and $\rho_R^B$ in
layer $\mathcal{A}$ and $\mathcal{B}$ as a function of transmission probability $\beta$, where
(a), (b) and (c) represent the cases of average degree $\langle k_{ab}\rangle=1$,
$\langle k_{ab}\rangle=3$, and $\langle k_{ab}\rangle=5$,
respectively. Based on the peaks of $\Delta$ in Fig. \ref{Fig:kab}(b), the red and green
dash lines indicate the first and second critical point, respectively. The symbols show the
simulated results and the solid lines are the corresponding theoretical results. All the parameters are
set as Fig. \ref{Fig:kab}. } \label{Fig:rhoAB}
\end{figure}

\begin{figure}
\centering
\includegraphics[width=0.8\linewidth]{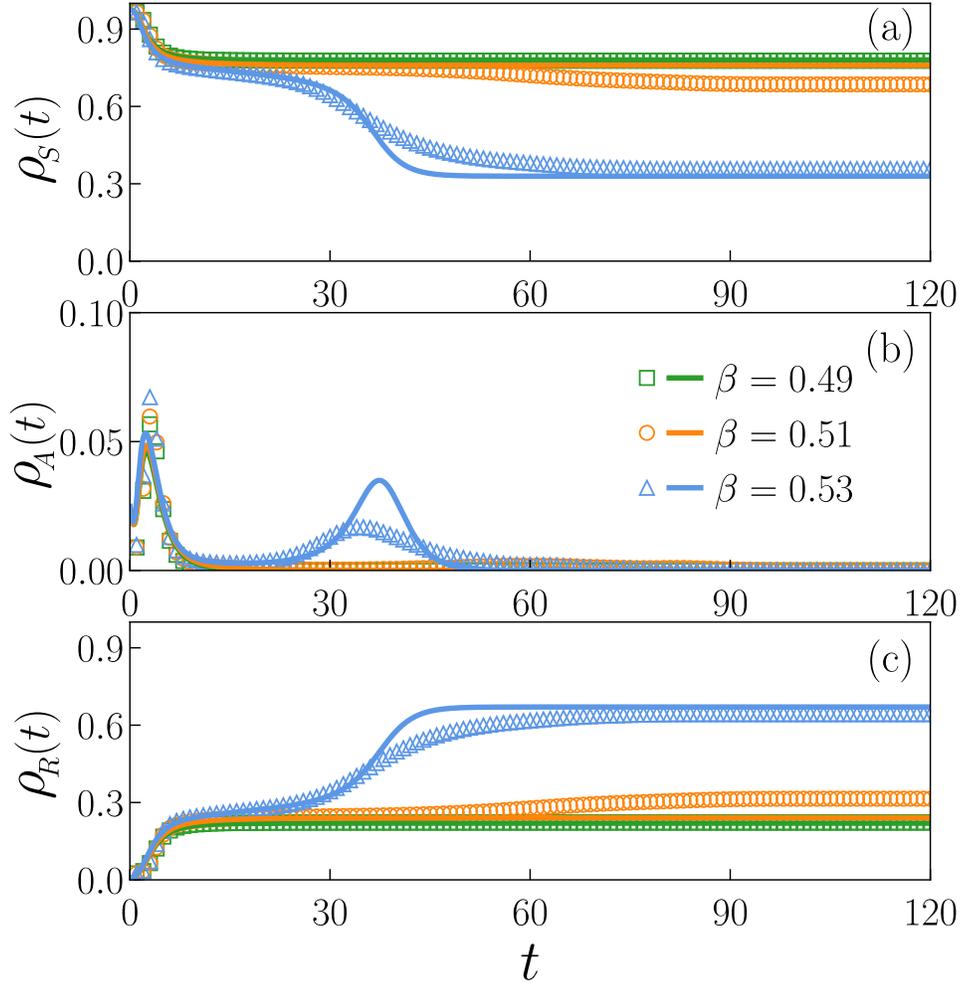}
\caption{(Color online). Average densities of (a) susceptible $\rho_S(t)$, (b) accepted $\rho_A(t)$,
and (c) recovered $\rho_R(t)$ nodes versus time $t$ with different transmission probability
$\beta$. Squares, circles and up triangles represent $\beta=0.49$, $0.51$ and $0.53$, respectively,
which indicate the spreading patterns below, at and above the second transition in Fig.
\ref{Fig:kab}(a) with $\langle k_{ab}\rangle=1$. The symbols show the simulated
results and the lines are the corresponding theoretical results. All the results are averaged over
$100$ independent realizations and the parameters are the same as Fig.
\ref{Fig:kab}. } \label{Fig:timeseries}
\end{figure}
To gather a deeper understanding the double transition phenomenon, in Fig. \ref{Fig:rhoAB} we
also measure the densities of final recovered nodes $\rho_R^A$ and $\rho_R^B$ in the layer
$\mathcal{A}$ and $\mathcal{B}$ as a function of transmission probability $\beta$, where Fig.
\ref{Fig:rhoAB}(a), (b) and (c) report the cases of average degree $\langle k_{ab}\rangle=1$, $\langle k_{ab}\rangle=3$, and
$\langle k_{ab}\rangle=5$, respectively. Based on the peaks of $\Delta$ in Fig. \ref{Fig:kab}(b), the red and
green dash lines indicate the first and second critical point $\beta_c^1$ and $\beta_c^2$,
respectively. Comparing with Fig. \ref{Fig:rhoAB}(a), (b) and (c), it is visible to observe that the
first threshold $\beta_c^1$ is the same, indicating that the first critical point $\beta_c^1$
corresponds to the spreading threshold in layer $\mathcal{A}$. With the increase of $\beta$, more
and more individuals have accepted the information in layer $\mathcal{A}$ and more information
has been spread to layer $\mathcal{B}$. When the $\beta$ closes to the second threshold
$\beta_c^2$, the system undergoes an abrupt transition. For a very strong coupling (see Fig.
\ref{Fig:rhoAB}(c)), the two critical points merge into one, which shows a discontinues
transition as the traditional threshold model \cite{Dodds:2004,Wang:2015}. In addition, from
Fig. \ref{Fig:rhoAB}(a) and (b), it is found that when $\beta_c^1<\beta<\beta_c^2$,
the density of recovered nodes in layer $\mathcal{B}$ is not zero, indicating the information has
been spread to a small fraction individuals in layer $\mathcal{B}$ but these small part accepted
individuals are unable to trigger an outbreak of the information.

To better understand the spreading behavior around the second threshold $\beta_c^2$, we study the evolution of the
nodes densities of susceptible $\rho_S(t)$, accepted $\rho_A(t)$, and recovered $\rho_R(t)$ in Fig. \ref{Fig:timeseries}, respectively.
The green, yellow and blue symbols and lines represent the spreading cases of below, at and above the second critical point
$\beta_c^2$, respectively. It is apparent to observe that when $\beta>\beta_c^2$ (see the blue symbols and lines), $\rho_A(t)$ shows
two peaks in Fig. \ref{Fig:timeseries}(b) and $\rho_R(t)$ increases dramatically at the final stage in Fig.
\ref{Fig:timeseries}(c), implying that the system undergoes a second outbreak.

\subsection{The effects of the adoption threshold $T_A$ and $T_B$ }
\begin{figure}
\centering
\includegraphics[width=0.8\linewidth]{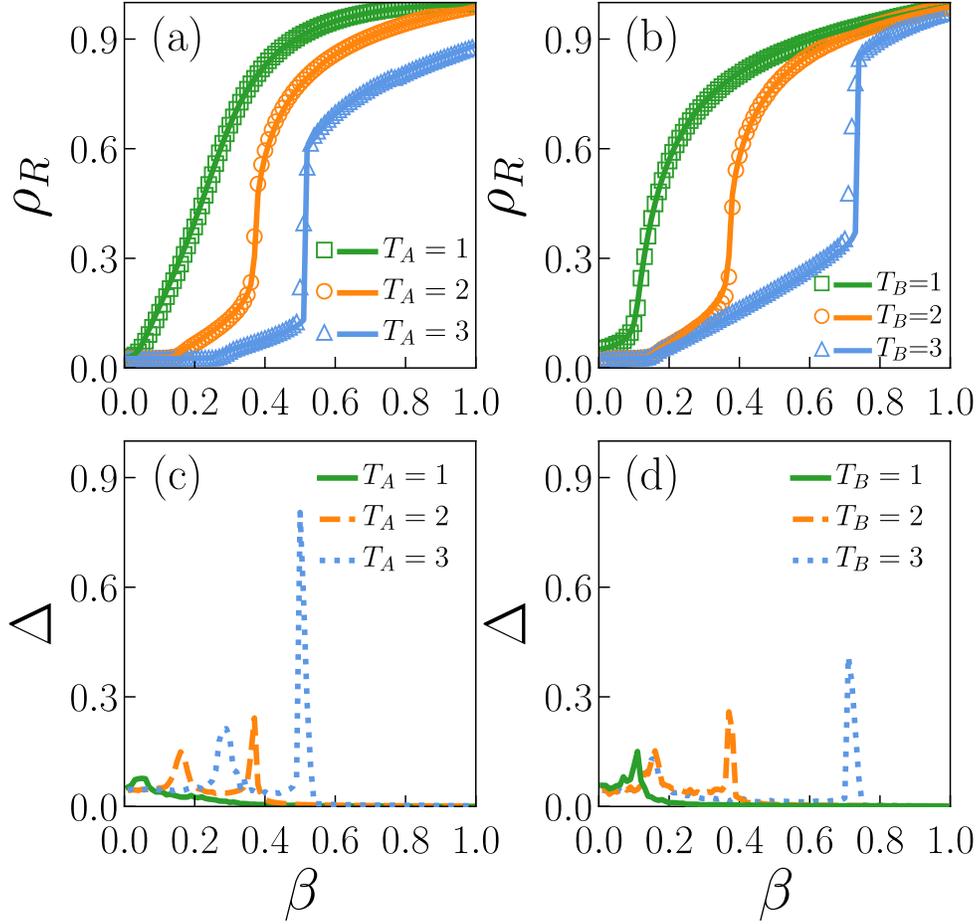}
\caption{(Color online). The effects of the adoption threshold $T_A$ and $T_B$ on double
transition. (a) and (b) show the dependence of the final recovered density $\rho_R$ on the
transmission probability $\beta$ with different $T_A$ and $T_B$, respectively. (c) and (d) plot the
corresponding variability $\Delta$ in the case of (a) and (b), respectively. The green, yellow and
blue symbols and lines represent $T_A=1$, $2$, $3$ in (a)(c) and $T_B=1$,
$2$, $3$ in (b)(d), respectively, where the symbols represent the simulated results and the lines
are the corresponding theoretical results in (a) and (b) from Eqs. (\ref{eq:25}) and (\ref{eq:26}).
The parameters are set as $T_B=2$ in (a) and $T_A=2$ in (b). The other ones are $N_a=N_b=10\,000$,
$\mu=1.0$, $\beta_{ab}=0.5$, $\langle k_{ab}\rangle=2$, $\rho_0=0.05$, $P_A(k_a)\sim k_a^{-2.1}$, $\langle k_a\rangle=6$,
$\langle k_b\rangle=6$.} \label{Fig:TaTb}
\end{figure}
In general, the adoption threshold of the individuals will influence the phase transition on the spreading dynamics\cite{Dodds:2004,Wang:2015}.
In this sense, we next study the effects of the adoption threshold $T_A$ and $T_B$ on the double transition.
Fig. \ref{Fig:TaTb}(a) and (b) show the dependence of the final recovered density $\rho_R$ on the
transmission probability $\beta$ with typical $T_A$ and $T_B$, respectively. As is shown in Fig.
\ref{Fig:TaTb}(a), when $T_A=1$, the phenomenon of double transition do not occur in the
spreading process. The corresponding variability $\Delta$ clearly confirms this point in Fig.
\ref{Fig:TaTb}(c). When $T_A=2$ and $T_A=3$, it is observed that the double
transition emerge with the increasing of $\beta$. The corresponding variability $\Delta$ shows two
peaks in Fig. \ref{Fig:TaTb}(c). Similarly, in Fig. \ref{Fig:TaTb}(b) and (d), we plot the final
recovered density $\rho_R$ and the corresponding variability $\Delta$ as a function of transmission
probability $\beta$ with different $T_B$, respectively. The results are similar to the case in Fig.
\ref{Fig:TaTb}(a) and (c). It is obvious to know that increasing the adoption threshold impedes
individuals from accepting the information. A larger value of adoption threshold means that
the individual will accept the information only it receives the information more times from distinct neighbors.
As a result, the individuals easily accept the
information when the adoption threshold is small (i.e., $T_A=1$ or $T_B=1$). Particularly, when
$T_A=1$, the information is spread fast in layer $\mathcal{A}$ and then the individuals in layer
$\mathcal{B}$ know the information quickly. Thus we observe a macroscopic outbreak at the
same time. While for the case of $T_B=1$, the individuals in layer $\mathcal{B}$ will accept the
information once they received it one time. In this case, the information in layer $\mathcal{A}$
can spill into layer $\mathcal{B}$ easily and it is equivalent to a relatively strong
interaction between the two layers, where the spreading process shows a synchronous outbreak behavior. Therefore,
the double transition disappears in this situation.

\subsection{Influence of network structure}
\begin{figure}
\centering
\includegraphics[width=0.8\linewidth]{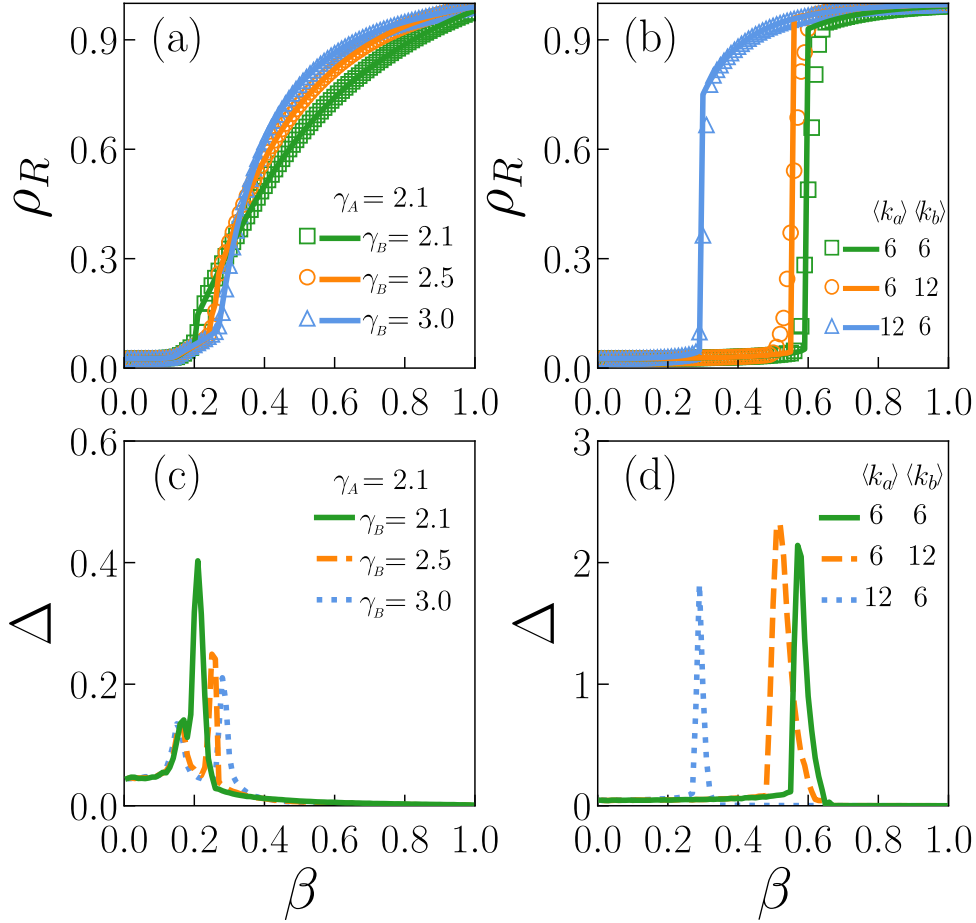}
\caption{(Color online). The influence of network structure on double transition. (a) and (c)
show $\rho_R$ and $\Delta$ versus $\beta$ with different degree exponent $\gamma_B$ in coupled
SF-SF networks, respectively. (b) and (d) show $\rho_R$ and $\Delta$ versus $\beta$ with different
$\langle k_a\rangle$ and $\langle k_b\rangle$ in coupled ER-ER networks, respectively. The degree exponent $\gamma_A=2.1$ is
fixed in (a) and (c) and the other parameters are set as $\langle k_{ab}\rangle=2$, $T_A=T_B=2$,
$N_a=N_b=10\,000$, $\mu=1.0$, $\beta_{ab}=0.5$, $\rho_0=0.05$.
} \label{Fig:sf}
\end{figure}

\begin{figure}
\centering
\includegraphics[width=0.8\linewidth]{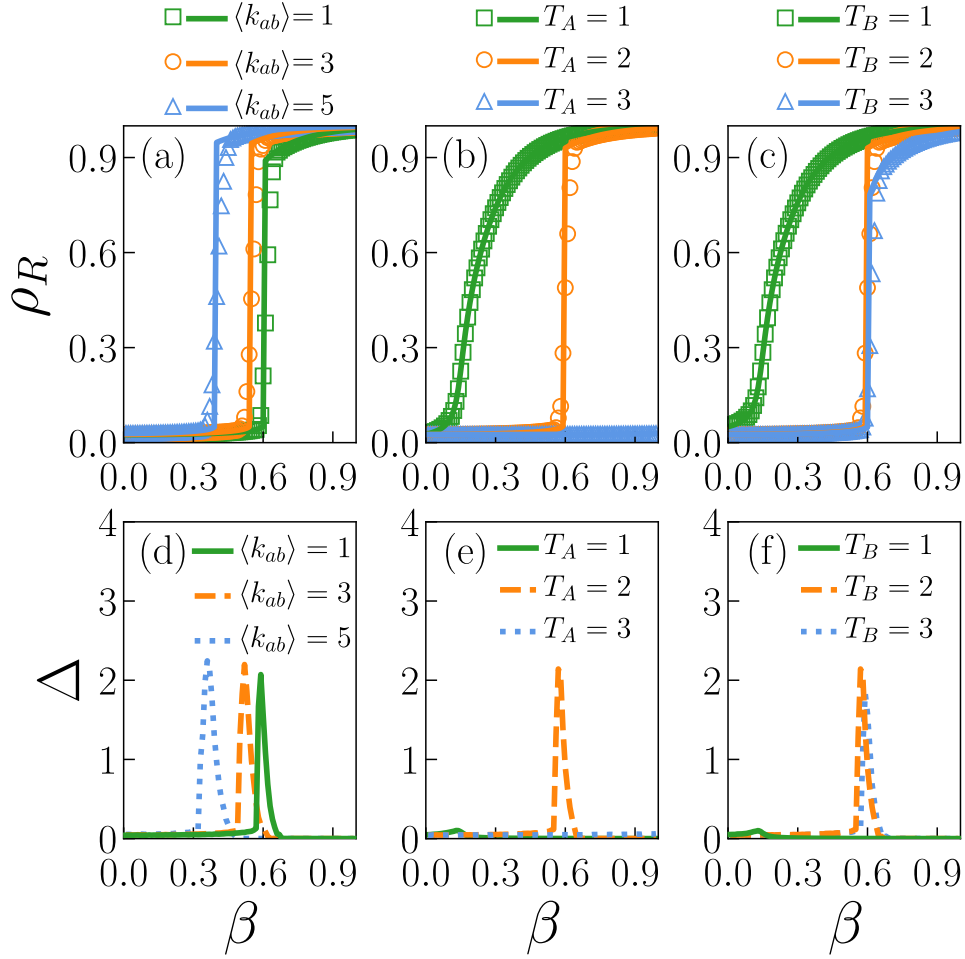}
\caption{(Color online). The double transition disappears on coupled ER-ER networks.
The dependence of the final recovered density $\rho_R$ on the transmission
probability $\beta$ with different (a)$\langle k_{ab}\rangle$, (b) $T_A$ and (c) $T_B$, respectively.
(d) (e) and (f) plot the corresponding variability $\Delta$ in the case of (a) (b) and (c), respectively.
The parameters are set as $T_A=T_B=2$ in (a)(d); $T_B=2$ in (b)(e); $T_A=2$ in (c)(f), respectively.
The other ones are $\langle k_a\rangle=\langle k_b\rangle=6$, $\langle k_{ab}\rangle=2$, $N_a=N_b=10\,000$,
$\mu=1.0$, $\beta_{ab}=0.5$, $\rho_0=0.05$.
}
\label{Fig:er}
\end{figure}
One more key question is how the network topology affects the phenomenon of the double transition. To answer
this question, we consider the influence of degree distribution of coupled SF-SF and ER-ER networks. Notice that
the coupled SF-SF network is generated with the power-law degree
distribution $P_A(k_a)\sim k_a^{-\gamma_A}$ and $P_B(k_b)\sim k_b^{-\gamma_B}$ in layer
$\mathcal{A}$ and $\mathcal{B}$, respectively, where $\gamma_A$ and $\gamma_B$ are the degree exponents.
The smaller of the degree exponent is, the stronger of the heterogeneity of network structure will be.
For fixed $\gamma_A=2.1$, Fig. \ref{Fig:sf} (a) and (c) show $\rho_R$
and $\Delta$ versus $\beta$ with different degree exponent $\gamma_B$ in coupled SF-SF networks, respectively.
It is found that when the $\gamma_B$ closes to $\gamma_A$, the phenomenon of the
double transition is not prominent any more. As the difference of the heterogeneity
in degree distribution between the two layers is not distinctive, the spreading speed in layer $\mathcal{A}$ and $\mathcal{B}$ are
comparative. In this case, it is easy to observe the synchronous outbreak behavior between layer $\mathcal{A}$ and $\mathcal{B}$.
In fact, the result can be qualitatively explained as follows\cite{Wang:2015}: From our model, we know that hubs
accept the information with more larger probability. With the increasing
of network heterogeneity in layer $\mathcal{B}$, the network has
a large number of nodes with very small degrees and
more nodes with large degrees. At the beginning, the hubs facilitate the information spreading as they are
more likely to receive the information from layer $\mathcal{A}$. After that, a large number of nodes in layer
$\mathcal{B}$ with very small degrees will
accept the information, resulting a similar behavior as layer $\mathcal{A}$ in the spreading process.

To deeply understand this point, we investigate a specific case with the same degree distribution
in coupled ER-ER networks. As is shown in Fig. \ref{Fig:sf}(b) and (d), both the curves of
$\rho_R$ and $\Delta$ indicate that the double transition disappears with different $\langle k_a\rangle$ and
$\langle k_b \rangle$. What is more, for different cases of $\langle k_{ab}\rangle$, $T_A$ and $T_B$, the disappearance of the
double transition is also found in Fig. \ref{Fig:er}. In an ER network,
the individuals are more likely to accept or not accept
the information synchronously, which result in a discontinuous transition \cite{Wang:2015}. These results
confirm again that the heterogeneity of degree distribution in each layer is very helpful for the appearance of
the double transition.
\section{Conclusions}

In recent years, researchers found that under certain conditions, there exists a double
transition in the infected fraction versus the transmission probability on the epidemic spreading
process. However, it is not clear whether it exists in the information spreading dynamics as the
information spreading carries its special features, such as the effects of multiple channels,
memory effects and non-redundant contacts etc. By combining these key factors in the
information spreading dynamics, we indeed find the double transition in the phase diagram.
These special features play a crucial role on the appearance of the double transition.

In summary, we have proposed a SAR model to describe the information spreading process on a two-layered network,
where we emphasize the effects of multiple channels, memory and non-redundant contacts. Our simulation results show that
there is a double transition in the phase diagram. Moreover, we find
that such a phenomenon originates from two outbreaks between the two networks, which is a
distinctive feature of a multilayer network of interactions. Further, we reveal that the double
transition are driven by a weak coupling condition between the two layers, a
large adoption threshold and the difference of the degree distributions betwen the two
networks. An edge-based compartmental theory is developed which fully
explains all numerical results. Our findings may be helpful
for understanding the secondary outbreaks of the information in our life.

\section{Acknowledgements}
This work was partially supported by the NNSF of China under Grant No. 11505114 and No.
10975099, the Program for Professor of Special Appointment (Orientational Scholar) at Shanghai
Institutions of Higher Learning under Grants No. QD2015016, and the Fundamental Research Funds
for the central Universities under Grants No. YJ201830.

\end{CJK*}

\section{References}

\begin{thebibliography}{}
\bibitem{Pastor-Satorras:2001}
R. Pastor-Satorras and A. Vespignani,
Phys. Rev. Lett. {\bf 86}, 3200 (2001).

\bibitem{Boguna:2002}
M. Boguna and R. Pastor-Satorras,
Phys. Rev. E {\bf 66}, 047104 (2002).

\bibitem{Ferreira:2012}
S. C. Ferreira, C. Castellano and R. Pastor-Satorras,
Phys. Rev. E {\bf 86}, 041125 (2012).

\bibitem{Boguna:2013}
M. Boguna, C. Castellano and R. Pastor-Satorras, Phys. Rev. Lett. {\bf 111}, 068701 (2013).

\bibitem{Parshani:2010}
R. Parshani, S. Carmi, and S. Havlin, Phys. Rev. Lett. {\bf 104},
258701 (2010).

\bibitem{Castellano:2010}
C. Castellano and R. Pastor-Satorras, Phys. Rev. Lett. {\bf 105},
218701 (2010).

\bibitem{Colizza:2007a}
V. Colizza, R. Pastor-Satorras and A. Vespignani,  Nature Phys. {\bf
3}, 276-282 (2007).

\bibitem{Colizza:2007b}
V. Colizza and A. Vespignani,
Phys. Rev. Lett. {\bf 99}, 148701 (2007).

\bibitem{Andrea:2008}
A. Baronchelli, M. Catanzaro and R. Pastor-Satorras, Phys. Rev. E {\bf 78}, 016111 (2008).

\bibitem{Liu:2009}
M. Tang, L. Liu and Z. Liu, Phys. Rev. E {\bf 79}, 016108 (2009).

\bibitem{Boccaletti:2014}
S. Boccaletti, G. Bianconi, R. Criado, C. I. del Genio, J. G\'{o}mez-Garde\~{n}es, M. Romance,
I. Sendi\~{n}a-Nadal, Z. Wang, and M. Zanin, Phys. Rep. \textbf{544}, 1 (2014).

\bibitem{Feng:2015}
L. Feng, C. P. Monterola, and Y. Hu, New J. Phys. \textbf{17}(6), 063025 (2015).

\bibitem{Sahneh:2013}
F. D. Sahneh, C. Scoglio, and F. N. Chowdhury, In 2013 American Control Conference (pp. 2307-2312). IEEE (2013).

\bibitem{Wang:2013}
H. Wang, Q. Li, G. D'Agostino, S. Havlin, H. E. Stanley, and P. Van Mieghem, Phys. Rev. E \textbf{88}(2), 022801 (2013).

\bibitem{Yagan:2013}
O. Yagan, D. Qian, J. Zhang, and D. Cochran, IEEE J. Sel. Areas Commun. \textbf{31}(6), 1038-1048 (2013).

\bibitem{Newman:2005}
M. E. Newman, Phys. Rev. Lett. \textbf{95}(10), 108701 (2005).
\bibitem{Marceau:2011}
V. Marceau, P. A. No\"{e}l, L. H\'{e}bert-Dufresne, A. Allard, and L. J. Dub\'{e},
Phys. Rev. E \textbf{84}, 026105 (2011).

\bibitem{Buono:2015}
C. Buono, and L. A. Braunstein, Europhy. Lett. \textbf{109}(2), 26001 (2015).

\bibitem{Buono:2014}
C. Buono, L. G. Alvarez-Zuzek, P. A. Macri, and L. A. Braunstein,
PloS One \textbf{9}(3), e92200 (2014).

\bibitem{Zhao:2014}
Y. Zhao, M. Zheng and Z. Liu, Chaos {\bf 24}, 043129 (2014).

\bibitem{Zheng:2017}
M. Zheng, M. Zhao, B. Min, and Z. Liu, Sci. Rep., \textbf{7}, 2424 (2017).

\bibitem{Zheng:2018}
M. Zheng, W. Wang, M. Tang, J. Zhou, S. Boccaletti, and Z. Liu, Chaos, Solitons \& Fractals, \textbf{107} (2018) 135-142.

\bibitem{Holme:2012}
P. Holme and J. Saramaki, Phys. Rep. {\bf 519}, 97-125 (2012).

\bibitem{Perra:2012}
N. Perra, B. Goncalves, R. Pastor-Satorras and A. Vespignani, Sci. Rep. {\bf 2}, 469 (2012).

\bibitem{Pastor:2015}
R. Pastor-Satorras, C. Castellano, P. V. Mieghem and A. Vespignani,
Rev. Mod. Phys. {\bf 87}, 925 (2015).

\bibitem{Barrat:2008}
A. Barrat, M. Barthelemy, and A. Vespignani, {\it Dynamical
Processes on Complex Networks} (Cambridge University
Press, Cambridge, England, 2008).

\bibitem{Dorogovtsev:2008}
S. N. Dorogovtsev, A. V. Goltsev, and J. F. F. Mendes, Rev.
Mod. Phys. {\bf 80}, 1275 (2008).

\bibitem{Wang:2017}
W. Wang, M. Tang, H. E. Stanley, and L. A. Braunstein,
Rep. Prog. Phys., \textbf{80}(3), 036603 (2017).


\bibitem{Simon:2014}
P. Colomer-de Simon , M. Bogu\~{n}\'{a} , Phys. Rev. X \textbf{4}, 041020 (2014).

\bibitem{Bhat:2017}
U. Bhat, M. Shrestha, L. H\'{e}bert-Dufresne, Phys. Rev. E \textbf{95}, 012314 (2017).

\bibitem{Allard:2017}
A. Allard, B. M. Althouse, S. V. Scarpino, and L. H\'{e}bert-Dufresne,
Proc. Natl. Acad. Sci. USA \textbf{114}(34), 8969-8973 (2017).

\bibitem{Min:2017}
B. Min, and M. S. Miguel, arXiv:1712.05059 (2017).

\bibitem{Dodds:2004}
P. S. Dodds, and D. J. Watts, Phys. Rev. Lett. \textbf{92} 218701 (2004).

\bibitem{Lu:2011}
L. L\"{u}, D.-B. Chen, T. Zhou, New J. Phys. {\bf13}, 123005 (2011).


\bibitem{Zheng:2013}
M. Zheng, L. L\"{u}, and M. Zhao, Phys. Rev. E \textbf{88}(1), 012818 (2013).

\bibitem{Wang:2015}
W. Wang, M. Tang, H. F. Zhang, and Y. C. Lai, Phys. Rev. E \textbf{92}, 012820 (2015).

\bibitem{Wu:2018}
J. Wu, M. Zheng, Z. K. Zhang, W. Wang, C. Gu, and Z. Liu,  Chaos \textbf{28}(3), 033113 (2018).

\bibitem{Brummitt:2012}
C. D. Brummitt, K. M. Lee, and K. I. Goh, Phys. Rev. E \textbf{85}(4), 045102 (2012).

\bibitem{Lee:2014}
K. M. Lee, C. D. Brummitt, and K. I. Goh, Phys. Rev. E \textbf{90}(6), 062816 (2014).

\bibitem{Min:2016}
B. Min, S. H. Gwak, N. Lee, and K. I. Goh, Sci. Rep. \textbf{6}, 21392 (2016).

\bibitem{Funk:2010}
S. Funk, and V. A. Jansen, Phys. Rev. E \textbf{81}, 036118 (2010).

\bibitem{Allard:2009}
A. Allard, P. A. No\"{e}l, L. J. Dub\'{e}, and B. Pourbohloul, Phys. Rev. E \textbf{79}, 036113 (2009).

\bibitem{Son:2012}
S. W. Son, G. Bizhani, C. Christensen, P. Grassberger, and M. Paczuski, Europhy. Lett. \textbf{97}, 16006 (2012).

\bibitem{Sanz:2012}
J. Sanz, C. Y. Xia, S. Meloni, and Y. Moreno, Phys. Rev. X \textbf{4}, 041005 (2014).

\bibitem{Souza:2009}
E. A. Leicht, and R. M. D'Souza, arXiv:0907.0894 (2009).

\bibitem{Hackett:2016}
A. Hackett, D. Cellai, S. G\'{o}mez, A. Arenas, and J. P. Gleeson, Phys. Rev. X \textbf{6}(2), 021002 (2016).


\bibitem{Dickison:2012}
M. Dickison, S. Havlin, and H. E. Stanley, Phys. Rev. E \textbf{85}, 066109 (2012).

\bibitem{Mendiola:2012}
A. Saumell-Mendiola, M. A. Serrano, and M. Bogu\~{n}\'{a}, Phys. Rev. E \textbf{86}, 026106 (2012).

\bibitem{Volz:2008}
E. Volz, J. Math. Biol \textbf{56}(3), 293-310 (2008).

\bibitem{Miller:2011}
J. C. Miller, J. Math. Biol. \textbf{62}(3), 349-358 (2011).

\bibitem{Shu:2016}
P. Shu, W. Wang, M. Tang, P. Zhao, and Y. C. Zhang, Chaos \textbf{26}(6), 063108 (2016).

\bibitem{Miller:2012}
J. C. Miller, A. C. Slim, and E. M. Volz, J. R. Soc. Interface \textbf{9}(70), 890-906 (2012).

\bibitem{Miller:2013a}
J. C. Miller, and  E. M. Volz, PloS One \textbf{8}(8), e69162 (2013).

\bibitem{Miller:2014}
J. C. Miller, PloS one \textbf{9}(7), e101421 (2014).

\bibitem{Catanzaro:2005}
M. Catanzaro, M. Bogu\~{n}\'{a}, and R. Pastor-Satorras, Phys. Rev. E {\bf 71}, 027103 (2005).

\bibitem{Albert:2002}
R. Albert, and A.-L. Barab\'asi, Rev. Mod. Phys. {\bf 74}, 47-97 (2002).

\bibitem{Crepey:2015}
P. Cr\'{e}pey, F. P. Alvarez, and M. Barth\'{e}lemy, Phys. Rev. E \textbf{73}, 046131 (2006).

\bibitem{Radicchi:2013}
F. Radicchi, and A.Arenas, Nat. Phys. \textbf{9}(11), 717-720 (2013).

\bibitem{Sahneh:2015}
F. D. Sahneh, C. Scoglio, and P. Van Mieghem, Phys. Rev. E \textbf{92}(4), 040801 (2015).




\end{thebibliography}
\end{document}